\documentclass[a4paper,british]{elsarticle}
\usepackage[T1]{fontenc}
\usepackage[latin9]{inputenc}
\usepackage{verbatim}
\usepackage{amsmath}
\usepackage{amssymb}
\usepackage{graphicx}

\makeatletter

\providecommand{\tabularnewline}{\\}

\def\e{\mathrm{e}}

\def\RE{\mathit{Re}}
\def\vec#1{\mbox{\boldmath $\mathit{#1}$}}
\def\tilde#1{\mbox{\boldmath $\mathrm{#1}$}}
\def\underline#1{\mbox{\boldmath $\mathsf{#1}$}}

\makeatother

\usepackage{babel}

\makeatother

\usepackage{babel}
\begin{document}

\title{Capacitance matrix technique for avoiding spurious eigenmodes in
the solution of hydrodynamic stability problems by Chebyshev collocation
method }

\author{Jonathan Hagan and J\={a}nis Priede}

\address{Applied Mathematics Research Centre, Coventry University, Coventry,
CV1 5FB, UK}
\begin{abstract}
We present a simple technique for avoiding physically spurious eigenmodes
that often occur in the solution of hydrodynamic stability problems
by the Chebyshev collocation method. The method is demonstrated on
the solution of the Orr-Sommerfeld equation for plane Poiseuille flow.
Following the standard approach, the original fourth order differential
equation is factorised into two second-order equations using a vorticity-type
auxiliary variable with unknown boundary values which are then eliminated
by a capacitance matrix approach. However the elimination is constrained
by the conservation of the structure of matrix eigenvalue problem,
it can be done in two basically different ways. A straightforward
application of the method results in a couple of physically spurious
eigenvalues which are either huge or close to zero depending on the
way the vorticity boundary conditions are eliminated. The zero eigenvalues
can be shifted to any prescribed value and thus removed by a slight
modification of the second approach.\end{abstract}
\maketitle

\section{Introduction}

Spectral methods are known to achieve exponential convergence rate
\citep{Canuto-etal}, which makes them particularly useful for solving
numerically demanding differential eigenvalue problems which arise
in hydrodynamic stability analysis \citep{Orszag-71}. Unfortunately,
besides providing accurate and efficient solutions for a certain number
of leading eigenvalues, spectral methods often produce physically
spurious unstable modes, which cannot be removed by increasing the
numerical resolution \citep{Gottlieb-Orszag-77}. For detailed discussion
of these modes we refer to Boyd \citep{Boyd-01}. Such physically
spurious eigenvalues can appear in all types of spectral methods including
Galerkin \citep{Zebib-84}, tau \citep{Dawkins-Dunbar-Douglass-98}
and collocation approximations \citep{Canuto-etal}, unless some kind
of \emph{ad hoc} approach is applied to avoid them. In the Galerkin
method, spurious eigenvalues can be removed by using the basis functions
also as the test functions instead of separate Chebyshev polynomials
\citep{Zebib-87}. A number of approaches avoiding spurious eigenvalues
have also been found for the tau method \citep{Gardner-Trogdon-Douglass-89,McFadden-Murray-Boisvert-90,Lindsay-Ogden-92}.
The same can be achieved also for the collocation (or pseudospectral)
method by using two distinct interpolating polynomials \citep{Huang-Sloan-94}.
Following the approach of McFadden \emph{et al}. \citep{McFadden-Murray-Boisvert-90}
for the tau method, Huang and Sloan \citep{Huang-Sloan-94} use a
Lagrange interpolating polynomial for second-order terms which is
by two orders lower than the Hermite interpolant used for other terms.
The choice of the latter polynomial depends on the particular combination
of the boundary conditions for the problem to be solved \citep[p. 493]{Weideman-Reddy-00}.

The objective of this paper is to present a simple method avoiding
spurious eigenmodes in the Chebyshev collocations method which uses
only the Lagrange interpolating polynomial applicable to general boundary
conditions. Our approach is based on the capacitance matrix technique
which is used to eliminate fictitious boundary conditions for a vorticity-type
auxiliary variable. The elimination can be performed in two basically
different ways which respectively produce a pair of infinite and zero
spurious eigenvalues. The latter can be shifted to any prescribed
value by a simple modification of the second approach. The main advantage
of our method is not only its simplicity but also applicability to
more general problems with complicated boundary conditions.

The paper is organised as follows. In the next section we introduce
the Orr-Sommerfeld problem for plane Poiseuille flow, which is a standard
test case for this type of method. Section \ref{sec:Cheb-col} presents
the basics of the Chebyshev collocation method that we use. The elimination
of the vorticity boundary conditions, which constitutes the basis
of our method, is performed in Sec. \ref{sec:Method}. Section \ref{sec:Num-res}
contains numerical results for the Orr-Sommerfeld problem of plane
Poiseuille flow. The paper is concluded by a summary of results in
Sec. \ref{sec:Sum}.

\section{Hydrodynamic stability problem}

The method will be developed by considering the standard hydrodynamic
stability problem of plane Poiseuille flow of an incompressible liquid
with density $\rho$ and kinematic viscosity $\nu$ driven by a constant
pressure gradient $\vec{\nabla}p_{0}=-\vec{e}_{x}P_{0}$ in the gap
between two parallel walls located $z=\pm h$ in the Cartesian system
of coordinates with the $x$ and $z$ axes directed streamwise and
transverse to the walls, respectively. The velocity distribution $\vec{v}(\vec{r},t)$
is governed by the Navier-Stokes equation 
\begin{equation}
\partial_{t}\vec{v}+(\vec{v}\cdot\vec{\nabla})\vec{v}=-\rho^{-1}\vec{\nabla}p+\nu\vec{\nabla}^{2}\vec{v}\label{eq:NS}
\end{equation}
 and subject to the incompressiblity constraint $\vec{\nabla}\cdot\vec{v}=0.$
Subsequently, all variables are non-dimensionalised by using $h$
and $h^{2}/\nu$ as the length and time scales, respectively. Note
that instead of the commonly used maximum flow velocity, we employ
the viscous diffusion speed $\nu/h$ as the characteristic velocity.
This non-standard choice will allow us to test our numerical method
against the analytical eigenvalue solution for a quiescent liquid.

The problem above admits a rectilinear base flow $\vec{v}_{0}(z)=\RE\bar{u}(z)\vec{e}_{x},$
where $\bar{u}(z)=1-z^{2}$ is the parabolic velocity profile and
$\RE=U_{0}h/\nu$ is the Reynolds number defined in terms of the maximum
flow velocity $U_{0}=2P_{0}h^{2}/\rho\nu.$ Stability of this base
flow is analysed with respect to small-amplitude perturbations $\vec{v}_{1}(\vec{r},t)$
by searching the velocity as $\vec{v}=\vec{v}_{0}+\vec{v}_{1}$. Since
the base flow is invariant in both $t$ and $\vec{x}=(x,y),$ perturbation
can be sought as a Fourier mode
\begin{equation}
\vec{v}_{1}(\vec{r},t)=\vec{\hat{v}}(z)\e^{\lambda t+\mathrm{i}\vec{k}\cdot\vec{x}}+\mbox{c.c.},\label{eq:v1}
\end{equation}
 defined by a complex amplitude distribution $\vec{\hat{v}}(z)$,
temporal growth rate $\lambda$ and the wave vector $\vec{k}=(\alpha,\beta).$
The incompressiblity constraint, which takes the form $\vec{D}\cdot\vec{\hat{v}}=0,$
where $\vec{D}\equiv\vec{e}_{z}\frac{d\,}{dz}+\mathrm{i}\vec{k}$ is a spectral
counterpart of the nabla operator, is satisfied by expressing the
component of the velocity perturbation in the direction of the wave
vector as $\hat{u}_{\shortparallel}=\vec{e}_{\shortparallel}\cdot\vec{\hat{v}}=\mathrm{i} k^{-1}\hat{w}',$
where $\vec{e}_{\shortparallel}=\vec{k}/k$ and $k=|\vec{k}|.$ Taking
the \emph{curl} of the linearised counterpart of Eq. (\ref{eq:NS})
to eliminate the pressure gradient and then projecting it onto $\vec{e}_{z}\times\vec{e}_{\shortparallel},$
after some transformations we obtain the Orr-Sommerfeld equation 
\begin{equation}
\lambda\vec{D}^{2}\hat{w}=\vec{D}^{4}\hat{w}+\mathrm{i}\alpha\RE(\bar{u}''-\bar{u}\vec{D}^{2})\hat{w},\label{eq:OS}
\end{equation}
 which is written in a non-standard form corresponding to our choice
of the characteristic velocity. Note that the Reynolds number appears
in this equation as a factor at the convective term rather than its
reciprocal at the viscous term as in the standard form. As a result,
the growth rate $\lambda$ differs by a factor $\RE$ from its standard
definition. The same difference, in principle, applies also to the
velocity perturbation amplitude which, however, is not important as
long as only the linear stability is concerned. In this form, Eq.
(\ref{eq:OS}) admits a regular analytical solution at $\RE=0,$ which
is used as a benchmark for the numerical solution in Sec. \ref{sec:Num-res}.

The no-slip and impermeability boundary conditions require 
\begin{equation}
\hat{w}=\hat{w}'=0\quad\mbox{at}\quad z=\pm1.\label{eq:bc}
\end{equation}
 Because three control parameters $\RE$ and $(\alpha,\beta)$ appear
in Eq. (\ref{eq:OS}) as only two combinations $\alpha\RE$ and $\alpha^{2}+\beta^{2},$
solutions for oblique modes with $\beta\not=0$ are equivalent to
the transverse ones with $\beta=0$ and a larger $\alpha$ and, thus,
a smaller $\RE$ which keep both parameter combinations constant \citep{Drazin-Reid-81}.
Therefore, it is sufficient to consider only the transverse perturbations
$(k=\alpha)$.

The first step in avoiding spurious eigenvalues in the discretizied
version of Eq. (\ref{eq:OS}) to be derived in the following section
is to represent Eq. (\ref{eq:OS}) as a system of two second-order
equations \citep{Gottlieb-Orszag-77} 
\begin{eqnarray}
\lambda\hat{\zeta} & = & \vec{D}^{2}\hat{\zeta}+\mathrm{i}\alpha\RE(\bar{u}''\hat{w}-\bar{u}\hat{\zeta}),\label{eq:OS-zeta}\\
\hat{\zeta} & = & \vec{D}^{2}\hat{w},\label{eq:zeta}
\end{eqnarray}
 where $\hat{\zeta}$ is a vorticity-type auxiliary variable which
has no explicit boundary conditions.

\section{\label{sec:Cheb-col}Chebyshev collocation method}

The problem is solved numerically using a collocation method with
$N+1$ Chebyshev-Gauss-Lobatto nodes 
\begin{equation}
z_{i}=\cos\left(i\pi/N\right),\quad i=0,\cdots,N.\label{zcol}
\end{equation}
 at which the discretizied solution $(\hat{w},\hat{\zeta})(z_{i})=(w_{i},\zeta_{i})=(\tilde{w},\tilde{\zeta})$
and its derivatives are sought. The latter are expressed in terms
of the former by using the so-called differentiation matrices, which
for the first and second derivatives are denoted by $D_{i,j}^{(1)}$
and $D_{i,j}^{(2)}$ with explicit expressions given in the Appendix.
Requiring Eqs. (\ref{eq:OS-zeta},\ref{eq:zeta}) to be satisfied
at the internal collocation points $0<i<N$ and the boundary conditions
(\ref{eq:bc}) at the boundary points $i=0,N,$ the following system
of $2N$ algebraic equations is obtained for the same number of unknowns
\begin{eqnarray}
\lambda\tilde{\zeta}_{0} & = & \underline{A}\tilde{\zeta}_{0}+\underline{B}\tilde{\zeta}_{1}+\tilde{g}_{0},\label{eq:OS-zeta0}\\
\tilde{\zeta}_{0} & = & \underline{A}\tilde{w}_{0},\label{eq:zeta0}\\
\tilde{0}_{1} & = & \underline{C}\tilde{w}_{0},\label{eq:bc0}
\end{eqnarray}
 where $\tilde{0}$ is the zero matrix and the subscripts $0$ and
$1$  denote the parts of the solution at the inner and boundary collocation
points, respectively; $\tilde{w}_{1}=\tilde{0}_{1}$ due to the first
boundary condition (\ref{eq:bc}) and 
\begin{equation}
g_{i}=\mathrm{i}\alpha\RE(\bar{u}_{i}''w_{i}-\bar{u}_{i}\zeta_{i}).\label{eq:f}
\end{equation}
 The matrices 
\begin{eqnarray}
A_{i,j} & = & (\vec{D}^{2})_{i,j},\quad0<(i,j)<N,\label{eq:A}\\
B_{i,j} & = & (\vec{D}^{2})_{i,j},\quad0<i<N,j=0,N,\label{eq:B}
\end{eqnarray}
 represent the parts of the collocation approximation of the operator
\begin{equation}
(\vec{D}^{2})_{i,j}=D_{i,j}^{(2)}-\alpha^{2}I_{i,j}\label{eq:D}
\end{equation}
 using the inner and boundary points, respectively; $I_{i,j}$ is
the unity matrix. Equation (\ref{eq:bc0}) is a discretizied version
of the second boundary condition (\ref{eq:bc}) imposed on $\hat{w}'$
which is defined by the matrix
\begin{equation}
C_{ij}=D_{i,j}^{(1)},\quad i=0,N;0<j<N.\label{eq:C}
\end{equation}

Our goal is to reduce Eqs. (\ref{eq:OS-zeta0}-\ref{eq:bc0}) to the
standard matrix eigenvalue problem for $\hat{w}_{0}.$ First, $\tilde{\zeta}_{0}$
is eliminated from Eq. (\ref{eq:OS-zeta0}) by using Eq. (\ref{eq:zeta0}),
which results in 
\begin{equation}
\lambda\underline{A}\tilde{w}_{0}=\underline{A}^{2}\tilde{w}_{0}+\underline{B}\tilde{\zeta}_{1}+\tilde{g}_{0}.\label{eq:OS-w0}
\end{equation}
 Next, we can use Eq. (\ref{eq:bc0}) to eliminate $\tilde{\zeta}_{1}$
from the equation above. This, as shown in the next section, can be
done in two basically different ways.

\section{\label{sec:Method}Elimination of the vorticity boundary values}

In order to eliminate $\tilde{\zeta}_{1}$ from Eq. (\ref{eq:OS-w0})
using Eq. (\ref{eq:bc0}) we employ a modified capacitance (or influence)
matrix method. For the basics of this method, see \citep[p. 178]{Peyret-02}
and references therein. Modifications to the method are due to the
structure of the matrix eigenvalue problem which needs to be conserved
in the elimination process. The general capacitance matrix approach
suggests to express $\tilde{w}_{0}$ from Eq. (\ref{eq:OS-w0}) and
then to substitute it into Eq. (\ref{eq:bc0}), which then would result
in a system of linear equations for $\tilde{\zeta}_{1}.$ However,
as noted above, the elimination procedure must be linear in $\lambda$
for the eigenvalue problem structure to be conserved. It means that
$\tilde{w}_{0}$ can be expressed either from the right or left hand
side of Eq. (\ref{eq:OS-w0}) but not from the combination of both
sides as in the standard capacitance matrix approach for the time
stepping schemes.

Our first approach is to express $\tilde{w}_{0}$ from the r.h.s.
of Eq. (\ref{eq:OS-w0}) by inverting $\underline{A}^{2}$ and then
substituting it into the boundary condition (\ref{eq:bc0}), which
results in
\begin{equation}
\underline{C}\tilde{w}_{0}=\underline{C}\underline{A}^{-2}(\lambda\underline{A}\tilde{w}_{0}-\underline{B}\tilde{\zeta}_{1}-\tilde{g}_{0})=\tilde{0}_{1}.\label{eq:bc1}
\end{equation}
 Next, solving the equation above for
\begin{equation}
\tilde{\zeta}_{1}=(\underline{C}\underline{A}^{-2}\underline{B})^{-1}\underline{C}\underline{A}^{-2}(\lambda\underline{A}\tilde{w}_{0}-\tilde{g}_{0})\label{eq:zeta1-1}
\end{equation}
 and substituting it into Eq. (\ref{eq:OS-w0}), we obtain 
\begin{equation}
\lambda\underline{E}\underline{A}\tilde{w}_{0}=(\underline{A}^{2}+\underline{E}\underline{G})\tilde{w}_{0},\label{eq:OS1-gen}
\end{equation}
 where $\underline{G}\tilde{w}_{0}=\tilde{g}_{0}$ and 
\begin{equation}
\underline{E}=\underline{I}-\underline{B}(\underline{C}\underline{A}^{-2}\underline{B})^{-1}\underline{C}\underline{A}^{-2}.\label{eq:E}
\end{equation}
 It is important to notice that $\underline{E}\underline{B}=0\underline{B},$
which means that $\underline{E}$ is singular. Namely, it has a zero
eigenvalue of multiplicity two corresponding to two eigenvectors represented
by the columns of $\underline{B}.$ Representing Eq. (\ref{eq:OS1-gen})
as 
\begin{equation}
(\underline{A}^{2}+\underline{E}\underline{G})^{-1}\underline{E}\underline{A}\tilde{w}_{0}=\lambda^{-1}\tilde{w}_{0},\label{eq:OS1-std}
\end{equation}
 which is a standard eigenvalule problem for $\lambda^{-1},$ it is
obvious that zero eigenvalues of $\underline{E}$ result in two zero
eigenvalues $\lambda^{-1},$ which in turn correspond to infinite
eigenvalues $\lambda$ of the original Eq. (\ref{eq:OS1-gen}). A
way to avoid these spurious eigenvalues is described below.

Alternative approach to eliminate $\tilde{\zeta}_{1}$ is to express
$\lambda\tilde{w}_{0}$ from the l.h.s. of Eq. (\ref{eq:OS-w0}) by
inverting $\underline{A}$ and then substituting it into the boundary
condition (\ref{eq:bc0}), which results in
\begin{equation}
\lambda\underline{C}\tilde{w}_{0}=\underline{C}\underline{A}^{-1}(\underline{A}^{2}\tilde{w}_{0}+\underline{B}\tilde{\zeta}_{1}+\tilde{g}_{0})=\tilde{0}_{1}.\label{eq:bc2}
\end{equation}
 This equation can be solved for $\tilde{\zeta}_{1}$ similarly to
Eq. (\ref{eq:bc1}) as 
\begin{equation}
\tilde{\zeta}_{1}=(\underline{C}\underline{A}^{-1}\underline{B})^{-1}\underline{C}\underline{A}^{-1}(\underline{A}^{2}+\underline{G})\tilde{w}_{0},\label{eq:zeta1-2a}
\end{equation}
 which substituted in Eq. (\ref{eq:OS-w0}) leads to
\begin{equation}
\lambda\underline{A}\tilde{w}_{0}=\underline{F}(\underline{A}^{2}+\underline{G})\tilde{w}_{0},\label{eq:OS2-gen}
\end{equation}
 where the transformation matrix 
\begin{equation}
\underline{F}=\underline{I}-\underline{B}(\underline{C}\underline{A}^{-1}\underline{B})^{-1}\underline{C}\underline{A}^{-1}\label{eq:F}
\end{equation}
 is singular with two zero eigenvalues because it satisfies $\underline{F}\underline{B}=0\underline{B}$
similarly to the $\underline{E}$ considered above. In contrast to
the previous eigenvalue problem defined by Eq. (\ref{eq:OS1-gen}),
now the singular transformation matrix appears on the r.h.s. of Eq.
(\ref{eq:OS2-gen}) and thus it produces two zero rather than infinite
eigenvalues $\lambda.$ 

It is important to notice that zero eigenvalues represent an alternative
solution to Eq. (\ref{eq:bc2}), which can be satisfied not only by
the boundary condition (\ref{eq:bc0}) but also by $\lambda=0.$ Consequently,
these spurious eigenvalues can be shifted from zero to any value $\lambda_{0}$
by subtracting $\lambda_{0}\underline{C}\tilde{w}_{0}$ from both
sides of Eq. (\ref{eq:bc2}), which obviously does not affect the
true eigenmodes satisfying Eq. (\ref{eq:bc0}). As a result we obtain
\begin{equation}
\tilde{\zeta}_{1}=(\underline{C}\underline{A}^{-1}\underline{B})^{-1}\underline{C}\underline{A}^{-1}(\underline{A}(\underline{A}-\lambda_{0}\underline{I})+\underline{G})\tilde{w}_{0},\label{eq:zeta1-2b}
\end{equation}
 which substituted in Eq. (\ref{eq:OS-w0}) leads to the following
standard eigenvalue problem
\begin{equation}
\lambda\tilde{w}_{0}=(\underline{A}^{-1}\underline{F}(\underline{A}(\underline{A}-\lambda_{0}\underline{I})+\underline{G})+\lambda_{0}\underline{I})\tilde{w}_{0}.\label{eq:OS2-std}
\end{equation}
{} The complex matrix eigenvalue problems above are solved using the
LAPACK's ZGEEV routine \citep{LAPACK}.

\section{\label{sec:Num-res}Numerical results}

\begin{table}[ht]
\centering{}%
\begin{tabular}{c|c|c|c}
$N=8$  & $N=16$  & $N=32$  & Exact \tabularnewline
\hline 
$5.4285\times10^{16}$  & $\quad1.2597\times10^{17}$  & $\quad3.5670\times10^{16}$  & $-$\tabularnewline
$3.1699\times10^{16}$  & $-1.1600\times10^{17}$  & $-6.0842\times10^{18}$  & $-$\tabularnewline
$-9.3120595$  & $-9.3137399$  & $-9.3137399$  & $-9.3137399$\tabularnewline
$-20.709030$  & $-20.570571$  & $-20.570571$  & $-20.570571$\tabularnewline
$-39.297828$  & $-38.947806$  & $-38.947789$  & $-38.947789$\tabularnewline
$-66.057825$  & $-60.054233$  & $-60.055435$  & $-60.055435$\tabularnewline
$-73.670710$  & $-88.285123$  & $-88.299997$  & $-88.299997$\tabularnewline
 & $-119.43366$  & $-119.27480$  & $-119.27480$\tabularnewline
 & $-157.89593$  & $-157.38866$  & $-157.38866$\tabularnewline
 & $-199.64318$  & $-198.23234$  & $-198.23234$\tabularnewline
 & $-226.99053$  & $-246.21576$  & $-246.21576$\tabularnewline
 & $-384.38914$  & $-296.92876$  & $-296.92874$\tabularnewline
 & $-409.06660$  & $-354.78191$  & $-354.78176$\tabularnewline
 & $-961.90740$  & $-415.36266$  & $-415.36420$\tabularnewline
 & $-961.99676$  & $-483.07721$  & $-483.08684$\tabularnewline
 &  & $-553.58796$  & $-553.53879$\tabularnewline
 &  & $-711.16464$  & $-711.45255$\tabularnewline
 &  & $\vdots$  & $\vdots$\tabularnewline
\end{tabular}\caption{\label{tab:Re0} The eigenvalues found numerically by solving Eq.
(\ref{eq:OS1-std}) (method I) with various number of collocation
points $N$ for $\alpha=1$ and $\RE=0.$ The exact eigenvalues values
are the roots of the characteristic equation resulting from analytical
solution of Eq. (\ref{eq:OS}) for $\RE=0.$ }
\end{table}

In order to validate the approach developed above we start with $\RE=0$
for which Eq. (\ref{eq:OS}) can easily be solved analytically leading
to the characteristic equation 
\begin{equation}
\frac{\tanh(k)}{\tan(\sqrt{k^{2}-\lambda})}=\pm\left(\frac{k}{\sqrt{k^{2}-\lambda}}\right)^{\pm1},\label{eq:analytic}
\end{equation}
 which defines two branches of eigenvalues $\lambda$ for the even
and odd modes corresponding to the plus and minus signs in the above
expression. The eigenvalues resulting from Eq. (\ref{eq:OS1-std}),
which represents our first approach, are shown in Table \ref{tab:Re0}
for various numbers of collocation points along with the exact solution
defined by Eq. (\ref{eq:analytic}). As seen, this approach indeed
produces a couple of huge spurious eigenvalues, which are due to the
singularity of the transformation matrix $\underline{E}$ (\ref{eq:E})
pointed out above. At the same time, the numerical solution accurately
reproduces the leading eigenvalues of the exact solution. The accuracy,
however, decreases down the spectrum so that only a half of the exact
eigenvalues are reproduced by the numerical solution. The other half
are numerically spurious eigenvalues which are due to the discretization
of the problem \citep{Boyd-01}.

\begin{figure}
\begin{centering}
\includegraphics[width=0.5\textwidth]{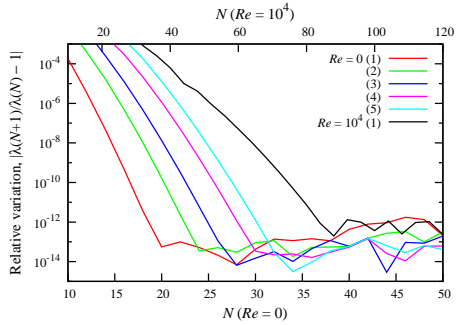} 
\par\end{centering}

\caption{\label{fig:k1}Relative variation of leading eigenvalues with the
number of collocation points $N$ for $\alpha=1$, $\RE=0$ and $\RE=10^{4}.$}
\end{figure}

Our second approach defined by Eq. (\ref{eq:OS2-gen}) produces exactly
the same eigenvalues as the first one for the given $N$ except for
the two spurious eigenvalues which are now machine-size zeros rather
than infinities. Using the modification of the second approach defined
by Eq. (\ref{eq:OS2-std}), these zero eigenvalues can be shifted
to any prescribed value $\lambda_{0}$ without affecting other eigenvalues.
Further we use $\lambda_{0}=4(N/4)^{4}$ which shifts the two physically
spurious eigenvalues to the region of numerically spurious eigenvalues
located in the lower part of spectrum. The variation of the five leading
eigenvalues with the number of collocation points $N$ plotted in
Fig. \ref{fig:k1} shows an exponential convergence rate characteristic
for the spectral numerical methods \citep{Canuto-etal}.

Next, we consider the solution of Eq. (\ref{eq:OS2-std}) for $\RE=10^{4}$
and $\alpha=1,$ which is a standard test case for the linear stability
analysis of plane Poiseuille flow. The leading eigenvalue for this
case is shown in table \ref{tab:Re10e4} in terms of the commonly
used phase velocity $c=-\mathrm{i}\lambda/\RE k.$ For $N\gtrsim60$ the solution
is seen to converge to the reference value obtained in \citep{Orszag-71}
using a tau method with $M\gtrsim30$ even Chebyshev polynomials.
As seen in Fig. \ref{fig:k1}, however the convergence rate for $\RE=10^{4}$
is somewhat slower than for $\RE=0,$ it is still exponential with
the final accuracy comparable to the previous case.

The number of collocation points can be reduced by a half by considering
even and odd modes separately as done in \citep{Orszag-71}. In our
case, this would require substitution of differentiation matrices
(\ref{eq:D1},\ref{eq:D2}) for general functions with their half-size
counterparts for even and odd functions, which, however, lies outside
the scope of this paper.

\begin{table}
\centering{}%
\begin{tabular}{c|c}
$N$  & $c$\tabularnewline
\hline 
16  & (0.23272286, 0.00922887)\tabularnewline
20  & (0.23814366, 0.00566303)\tabularnewline
24  & (0.23842504, 0.00282919)\tabularnewline
28  & (0.23735182, 0.00357013)\tabularnewline
32  & (0.23747200, 0.00372519)\tabularnewline
36  & (0.23752527, 0.00375400)\tabularnewline
40  & (0.23752494, 0.00373917)\tabularnewline
44  & (0.23752716, 0.00374012)\tabularnewline
48  & (0.23752633, 0.00373961)\tabularnewline
52  & (0.23752653, 0.00373969)\tabularnewline
56  & (0.23752648, 0.00373967)\tabularnewline
60  & (0.23752649, 0.00373967)\tabularnewline
64  & (0.23752649, 0.00373967)\tabularnewline
\end{tabular}\caption{\label{tab:Re10e4}Phase velocity $c=-\mathrm{i}\lambda/(\RE k)$ of the most
unstable mode depending on the number of collocation points $N$ for
$\alpha=1$ and $\RE=10^{4}.$ }
\end{table}

\section{\label{sec:Sum}Summary and conclusions}

We have developed a simple technique for avoiding physically spurious
eigenmodes in the solution of hydrodynamic stability problems by the
Chebyshev collocation method, which was demonstrated on the Orr-Sommerfeld
equation for plane Poiseuille flow. The method is based on the factorisation
of the original fourth order differential equation into two second-order
equations using a vorticity-type auxiliary variable which has no explicit
boundary conditions. The main element of the method is the elimination
of the vorticity boundary values by using a capacitance matrix approach
to obtain a standard matrix eigenvalue problem. Although the elimination
is constrained by of the structure of eigenvalue problem, it can be
still done in two basically different ways. Both approaches result
in couple of physically spurious eigenvalues, which are either huge
or close to zero depending on the way the vorticity boundary values
are eliminated. We showed that these spurious eigenvalues are due
to the double singularity of the transformation matrices which eliminate
the vorticity boundary conditions by multiplying either the stiffness
or mass matrices of the original generalised eigenvalue problem. By
a slight modification of the second approach, the zero eigenvalues
can be shifted to any prescribed value and thus moved to the region
of numerically spurious eigenvalues at the end of spectrum.

The main advantage of our method is not only its simplicity but also
its applicability to more general stability problems with complex
boundary conditions involving several variables. An example of such
a problem is that of 3D linear stability of MHD duct flow using a
non-standard vector stream function and vorticity formulation, which
results in the coupling of the stream function components through
the boundary conditions \citep{PAM10}. In this case neither Galerkin
nor collocation method with the \emph{ad hoc} approach of Huang and
Sloan \citep{Huang-Sloan-94} is applicable because no simple basis
functions satisfying the boundary conditions can be constructed.

\section*{Acknowledgement}

J.H. thanks the Mathematics and Control Engineering Department at
Coventry University for funding his studentship.

\section*{Appendix}

The differentiation matrices for the first and second derivatives
at Chebyshev-Gauss-Lobatto nodes (\ref{zcol}) are defined as follows
\citep[pp. 393--394]{Peyret-02}:

\begin{equation}
D_{i,j}^{(1)}=\left\{ \begin{array}{ll}
\frac{2N^{2}+1}{6} & j=0\\
\frac{c_{j}}{c_{i}}\frac{(-1)^{i+j}}{(z_{j}-z_{i})} & i\neq j\\
-\frac{z_{j}}{2(1-z_{j}^{2})} & 0<i=j<N\\
-\frac{2N^{2}+1}{6} & j=N
\end{array}\right.\label{eq:D1}
\end{equation}
 and 
\begin{equation}
D_{i,j}^{(2)}=(D_{i,j}^{(1)})^{2}=\left\{ \begin{array}{ll}
\frac{(-1)^{i+j}}{c_{j}}\frac{z_{i}^{2}+z_{i}z_{j}-2}{(1-z_{j}^{2})(z_{i}-z_{j})^{2}} & 0<i\neq j<N\\
-\frac{(N^{2}-1)(1-z_{i}^{2})+3}{3(1-z_{i}^{2})^{2}} & 0<i=j<N\\
\frac{2}{3}\frac{(-1)^{j}}{c_{j}}\frac{(2N^{2}+1)(1-z_{j})-6}{(1-z_{j})^{2}} & j\not=i=0\\
\frac{2}{3}\frac{(-1)^{j+N}}{c_{j}}\frac{(2N^{2}+1)(1+z_{j})-6}{(1+z_{j})^{2}} & j\not=i=N\\
\frac{N^{4}-1}{15} & i=j=0,N,
\end{array}\right.\label{eq:D2}
\end{equation}
 where $c_{i}=1$ for $0<i<N$ and $c_{i}=2$ for $i=0,N.$


\begin{thebibliography}{10}
\bibitem{LAPACK}E. Anderson, Z. Bai, C. Bischof, S. Blackford, J.
Demmel, J. Dongarra, J. Du Croz, A. Greenbaum, S. Hammarling, A. McKenney,
D. Sorensen, LAPACK Users' Guide, 3rd ed., SIAM, Philadelphia, 1999.

\bibitem{Boyd-01} J.P. Boyd, Chebyshev and Fourier Spectral Methods,
Dover, New York, 2001, pp. 139--142.

\bibitem{Canuto-etal} C. Canuto, M.Y. Hussaini, A. Quarteroni, T.A.
Zang, Spectral Methods in Fluid Dynamics, Springer, Berlin, 1988.

\bibitem{Dawkins-Dunbar-Douglass-98} P.T. Dawkins, S.R. Dunbar, R.W.
Douglass, The origin and nature of spurious eigenvalues in the spectral
tau method, J. Comput. Phys. \textbf{147} (1998) 441--462.

\bibitem{Dongarra-Straughan-Walker-96} J.J. Dongarra, B. Straughan,
D.W. Walker, Chebyshev tau-QZ algorithm methods for calculating spectra
of hydrodynamic stability problems, Appl. Num. Math. \textbf{22} (1996)
399--434.

\bibitem{Drazin-Reid-81}P.G. Drazin, W.H. Reid, Hydrodynamic Stability,
Cambridge, 1981, p. 141.

\bibitem{Gardner-Trogdon-Douglass-89} D.R. Gardner, S.A. Trogdon,
R.W. Douglass, A modified tau spectral method that eliminates spurious
eigenvalues, J. Comput. Phys\emph{.} \textbf{80} (1989) 137--167.

\bibitem{Gottlieb-Orszag-77} D. Gottlieb, S.A. Orszag, \textit{\emph{Numerical
Analysis of Spectral Methods: Theory and Applications}}, SIAM, Philadelphia,
1977, pp. 143--148.

\bibitem{Huang-Sloan-94} W. Huang, D.M. Sloan, The pseudospectral
method for solving differential eigenvalue problems, J. Comp. Phys.
\textbf{111} (1994) 399--409.

\bibitem{Lindsay-Ogden-92} K.A. Lindsay, R.R. Ogden, A practical
implementation of spectral methods resistant to the generation of
spurious eigenvalues, J. Numer. Methods Fluids \textbf{15} (1992)
1277--1294.%

\bibitem{McFadden-Murray-Boisvert-90} G.B. McFadden, B.T. Murray,
R.F. Boisvert, Elimination of spurious eigenvalues in the Chebyshev
tau spectral method, J. Comput. Phys. \textbf{91} (1990) 228--239.%

\bibitem{Orszag-71} S.A. Orszag, Accurate solution of the Orr-Sommerfeld
problem, J. Fluid Mech. \textbf{50} (1971) 689--703.

\bibitem{Peyret-02} R. Peyret, Spectral Methods for Incompressible
Viscous Flow, Springer, Berlin, 1982.

\bibitem{PAM10} J. Priede, S. Aleksandrova, S. Molokov, Linear stability
of Hunt's flow. J. Fluid Mech. \textbf{649} (2010) 115--134.%

\bibitem{Weideman-Reddy-00} J.A.C. Weideman and S.C. Reddy, A MATLAB
differentiation matrix suite, ACM Transactions on Mathematical Software
\textbf{26} (2000) 465--519.

\bibitem{Zebib-84} A. Zebib, A Chebyshev method for the solution
of boundary value problems, J. Comput. Phys. \textbf{53} (1984) 443--455.

\bibitem{Zebib-87} A. Zebib, Removal of spurious modes encountered
in solving stability problems by spectral methods, J. Comput. Phys.\emph{
}\textbf{70} (1987) 521--525. %
{} \end{thebibliography}
\end{document}